\documentclass[traditabstract]{aa}
\usepackage{natbib}

\bibpunct{(}{)}{;}{a}{}{,}
\usepackage{txfonts,graphicx}
\usepackage{endnotes,verbatim}
\usepackage[dvips]{color}
\usepackage{perpage,longtable}
\usepackage{subfigure,color}
\usepackage[caption=false]{subfig}
\newcommand\eg{{\it e.g.} }
\newcommand\vs{{\it vs. }}

\def\um{\ifmmode {\rm\mu m}\else {$\mu$m}\fi}
\MakePerPage{footnote}
\begin{document}
\title{Overdensities of 24\,$\mu$m Sources in the Vicinities of High-Redshift Radio Galaxies}
\author{Jack H. Mayo \inst{1} \and Jo\"{e}l Vernet\inst{2} \and Carlos De Breuck\inst{2} \and Audrey Galametz\inst{3}  \and Nick Seymour\inst{4} \and Daniel Stern\inst{5}}
\institute{Institute for Astronomy, University of Edinburgh, Blackford Hill, Edinburgh, EH9 3HJ. UK. [e-mail: {\tt jhm@roe.ac.uk}]
\and European Southern Observatory, Karl-Schwarzschild-Str. 2, D-85748 Garching, Germany.
\and INAF - Osservatorio di Roma, Via Frascati 33, I-00040, Monteporzio Catone, Italy.
\and CSIRO Astronomy \& Space Science, PO Box 76, Epping, NSW 1710, Australia.
\and Jet Propulsion Laboratory, California Institute of Technology, Pasadena, CA 91101, USA.}
\abstract{We present a statistical study of the environments of 63 high-redshift radio galaxies (HzRGs) between redshifts $1 \leq z \leq 5.2$, using the 24\,\um\, waveband of the MIPS instrument aboard the \emph{Spitzer Space Telescope}. Using a counts-in-cell analysis, a statistically significant source overdensity is found in $1.75\arcmin$ radius circular cells centred on the HzRGs when compared to reference fields. We report an average overdensity of $\delta\,(=\bar{N}_{targets} / \bar{N}_{reference})\,= 2.2\pm1.2$ at a flux density cut of $f_{24\um} =0.3$\,mJy.  This result implies that HzRGs are likely to lie in protoclusters of active and star-forming galaxies at high redshift. Over 95\% of our targeted HzRGs lie in higher than average density fields.  Further, 20 (32\%) of our selected fields are found to be overdense to at least a $3\sigma$ significance, of which 9 are newly identified protocluster candidates. We observe a weak correlation between redshift and 24\,\um\, source density, and discuss the populations being probed at different redshifts. In our uniformly selected sample, which was designed to cover two orders of magnitude in radio luminosity throughout $z=1-4$, we find that the 24\,\um\, source density does not depend on radio luminosity.  We also compare this result with recent work describing IRAC source overdensities around the same HzRGs and find correlations between the results.}
\keywords{clusters: observations $-$ galaxies: evolution $-$ galaxies: formation $-$ infrared: galaxies $-$ large scale structure}
\titlerunning{24\,\um\, Overdensities around HzRGs}
\authorrunning{J.H.Mayo et al.}
\maketitle
\date
\section{Introduction}
\label{motiv}
Observations of high-redshift radio galaxies (HzRGs: $z>1.0$ with a rest$-$frame 3GHz luminosity greater than $10^{26}\,$W$\,$Hz$^{-1}$) provide evidence that such galaxies are the progenitors of the current day giant elliptical (gE) and cluster-dominant (cD) galaxies \citep[see \eg][]{debreuck2002,seymour2007,debreuck10}. As such, we expect HzRGs to be signposts of high-density regions.  Studies have shown overdensities of Ly$\alpha$ emitters in HzRG environments between $2.2 \le z \le 4.1$ \citep[\eg][]{venemans2002,venemans2003,overzier2008}. 
 The formation of filamentary structures out to high redshifts \citep{croft2005,matsuda2005} and of Extremely Red Object (ERO) overdensities in high-redshift active galactic nuclei (AGN) fields have also been observed \citep[\eg][]{chapman2000,hall2001}. Overdensities of H$\alpha$ emitters \citep{kurk04b,tanaka11}, Lyman Break Galaxies \citep{miley04}, $BzK$-selected sources \citep{galametz09} and red sequence galaxies \citep{doherty10} have also been reported in the vicinities of HzRGs.
Large rotation measures within the HzRG environments \citep{carilli1997} are indicative of dense, X-ray emitting cluster atmospheres.  All of this evidence suggests HzRGs trace protoclusters, and indicate that protoclusters can be in place from early epochs (when the Universe was only $\sim 1\,-\,2$Gyr old).  These studies have been performed on limited samples of less than a dozen each; prior to this work, a systematic study of HzRG environments has not been undertaken. 

High-energy continuum emission from a central accretion disc in an AGN is believed to be re-processed by the dusty toroidal structure encompassing it, and re-radiated in the mid-IR \citep[$10-30\,\um$;][]{elvis1994}.  Since all AGN are believed to have such a torus, they will all have a mid-IR signature, and though there may be orientational effects, all AGN are thought to emit strongly in the rest-frame mid-IR.  However, mid-IR observations will not only detect AGN; starburst galaxies have strong polycyclic $-$ aromatic hydrocarbon (PAH) features, strongest at $\sim$3.3\,\um, $6.2$\,\um, $7.7$\,\um , $8.6$\,\um, $11.2$\,\um\, and $12.7$\,\um, attributed to vibrational modes of these complex molecules \citep{gillett1973}.  These attributes make mid-IR detection ideal for selection of star-forming galaxies at high redshifts as well. 

With blind targeting requiring large-area surveys, efforts to identify clusters can be both time and resource intensive.  If we can pre-select targets that are thought to preferentially reside within rich environments, we can quickly identify larger numbers of robust cluster candidates. Selecting HzRGs (which have been found to lie in overdense regions at other wavebands and at lower redshifts) and analysing their fields in the mid-IR allows us to probe the AGN and starburst populations simultaneously, without obvious selection effects.

The companion paper to this work, \citet[][ApJ, in press, and hereafter referred to as G12]{galametz12}, undertakes the analysis of 48 HzRG fields between redshift $1.2 \le z \le 3.0$ using the NASA's \emph{Spitzer Space Telescope} \citep{werner2004} Infra-Red Array Camera \citep[IRAC;][]{fazio04} to isolate high redshift sources.  Their work shows that HzRGs lie preferentially in medium-to-high density regimes when compared to a reference sample, though the colour criterion used in that work does not differentiate between passive and active galaxies. \\

Throughout this paper many references to the HzRG fields are made, they are addressed with the same IDs as the HzRGs, though typically we are discussing the environments and not the radio sources.  In this paper we refer to all significantly overdense regions as protoclusters or protocluster candidates. Though at low redshifts these structures may be gravitationally bound, protoclusters is used throughout for the sake of consistency.  A $\Lambda$CDM cosmology with $H_0 = 70 $\,km$ $s$^{-1} $Mpc$^{-1}$,  $\Omega_m = 0.3$ and $\Omega_{\Lambda} = 0.7$ is assumed throughout.
\section{Observations}
\subsection{HzRGs}
Our sample of HzRGs is taken from both flux$-$limited surveys such as 3C, 6CE, 7C and MRC for their unbiased radio properties, as well as ultrasteep radio spectra surveys \citep[\eg][]{debreuck2000} to cover the higher redshift ($z > 2$) range.  This sample is specifically designed to mitigate the luminosity-redshift correlation that affects flux-limited samples; the reader is referred to \citet[][Fig. 2]{debreuck10} for further details.

We present observations of the first statistically significant sample of HzRG fields to date. The 70 target \textit{Spitzer}-HzRG sample is described in more detail by \citet{seymour2007} where targets were preferentially selected based on their amount of supporting data (\textit{Hubble Space Telescope} and SCUBA observations) and existing guaranteed time observations (GTOs) with \textit{Spitzer}.  With no 24\,\um\, data (one source) and insufficient coverage (seven sources), our sample size is reduced to 62 targets.  The source \texttt{LBDS53W069}, although not meeting the HzRG radio luminosity selection criterion, has been included in this sample as it is a known radio galaxy with the relevant data available \citep{stern06}, making our HzRG sample 63 fields in total.

Our HzRG sample was observed at 24\,\um\, using the Multiband Imaging Photometer for \textit{Spitzer} \citep[MIPS;][]{rieke2004}.  The majority of observations (59 sources) have been undertaken as part of Program ID's 40093 (39 sources) and 3329 (20 sources) as part of \textit{Spitzer} proposal cycles 1 and 4 respectively.  Exposure times for the former 59 sources were determined based on 24\,\um\, background levels as determined with the Leopard software.  Data for three remaining sources were taken from much deeper GTO programs.

\subsection{Reference Fields}
The reference set against which our HzRG fields are analysed is that of the \textit{Spitzer} Wide-area InfraRed Extragalactic Survey \citep[SWIRE;][]{lonsdale2003}.  SWIRE, the widest area cryogenic \textit{Spitzer} legacy program, imaged 49 sq. deg. of the sky over six\footnote{ELAIS N1, ELAIS N2, ELAIS S1, Lockman, XMM$-$LSS, CDFS} high$-$latitude extragalactic fields in all seven available bands\footnote{IRAC: $3.6$\,\um, $4.5$\,\um, $5.8$\,\um, $8.0$\,\um. MIPS: $24$\,\um, $70$\,\um\,and $160$\,\um.} and detected over two million objects up to $z \sim 4$ \citep{rowan-robinson2005}.  The field choices for SWIRE required low Galactic cirrus emission, large contiguous areas and low foreground contamination such as bright stars and local galaxies.  The observing strategy for the SWIRE survey involved mapping of the fields such that each point is covered by four 30-second exposures with the IRAC instrument and 44 four-second exposures with the MIPS instrument.  The SWIRE observations were taken throughout 2004 with over 800 hours of observations.

After initial analysis of the SWIRE fields we opt to remove the ELAIS S1 field from comparison, due to its lower coverage compared to the remaining five SWIRE fields. By doing this, we manage to cut at a deeper flux density without jeopardising the signal to noise, whilst still allowing our sample to be large enough to mitigate cosmic variance.
\subsection{Reduction}
\label{section:reduction}
Data reduction was performed using MOPEX \citep{makovoz2005} as described by \citet{seymour2007}; basic calibrated data are drizzled to reach an increased sampling of $1\farcs25$ per pixel from the intrinsic pixel scale for MIPS 24\um\, frames of $2\farcs5$ per pixel\footnote{note that the SWIRE data is rescaled to $1\farcs2$ per pixel}.  
\section{Source Extraction and Photometry}
\subsection{Creation of Variance Maps}
Fully reduced data and corresponding coverage maps are used to produce accurate variance maps:  the pixel flux densities  are binned and plotted according to their coverage, Gaussian profiles are fit to the lower part of the resultant histograms (ignoring the positive skew due to sources), then finally RMS noise values are extracted for each coverage value.  Plotting of the root mean square (RMS) noise values \vs coverage (and using $\sigma^2 \propto 1/\cal{W}$, where $\sigma$ is RMS noise and $\cal{W}$ is the weight, equivalent to coverage) allows the coverage maps to be converted directly to \emph{absolute variance} maps.
\subsection{Detection and Extraction}
\label{det}
Source detection and photometry are undertaken using the SExtractor software \citep{bertin1996}. Parameters are set as to minimise false detections and to optimise the separate detection of sources in close proximity.  The parameters chosen are included in Appendix \ref{app2}.  The flux density from extended sources\footnote{area $>100$ pixels and stellarity  $<0.8$} is taken to be the Kron flux density, with the aperture correction to the flux density being calculated from the \emph{effective} Kron radius.  Elsewhere, the resultant flux density from point-like\footnote{area $<100$ pixels or stellarity $>0.8$} sources is taken to be that calculated within a $5\farcs25$ aperture radius.  Aperture corrections are calculated using the \emph{Spitzer} MIPS data handbook (see the MIPS data handbook\footnote{http://ssc.spitzer.caltech.edu/mips/dh/mipsdatahandbook3.3.1.pdf}, p.37).
\subsection{Flux Density and Region Cut}
Due to the dithering pattern adopted during the HzRG field observations for which the central region of each frame has a higher coverage, we restrain the present analysis to the central $1.75\arcmin$ radius circular fields centred on the HzRGs, this rejects areas in the frame with lower coverage.

To determine a suitable flux density cut to our data, we first calculate the $5\farcs25$ radius aperture errors for each frame.  This is done by placing random apertures upon the frame (within $1.75\arcmin$ radii circles), binning of source flux densities, and plotting the flux density distribution.  We again perform a Gaussian fit to the lower part of the distribution, so as to avoid the positive skew due to flux from sources, and calculate the Gaussian width; this width gives us the RMS noise for each frame (RMS values are included in Tables \ref{table1} and \ref{table2}).  Our RMS values calculated for the SWIRE fields are smaller than those of \citet{shupe08} by a factor of $\sim 1.35$; however, our results are consistent with their signal-to-noise measurements. For comparison, the average RMS noise for the HzRG fields is $23.0\,\mu$Jy, while for the SWIRE fields the average RMS noise is $29.6\,\mu$Jy.  We adopt a flux density cut of $f_{24\um} = 0.3$\,mJy, corresponding to a minimum $S/N \sim 5$ for the fields with the highest RMS, after aperture correction (both for our HzRGs and for SWIRE fields, this is limited by the SWIRE XMM-LSS fields). As will be shown in Sec. \ref{sect:analysis:RMS}, the HzRG RMS values are found to be independent of $24$\,\um\, source densities, which is indicative of a conservative cut free of questionable sources.
\subsection{Densities in $1.75\arcmin$ circular cells}
Contiguous $1.75\arcmin$ circular cells are placed intermittently (without overlap) within the five SWIRE fields, totalling some 9981 cells.  The 24\,\um\, source densities are then calculated in each of these cells, with an imposed flux density cut of $f_{24\um} = 0.3$\,mJy. Source densities in our HzRG fields are calculated by centreing the $1.75\arcmin$ radius apertures on the radio galaxy locations.  When undertaking a sources-in-cell count for our HzRG fields, the HzRG itself is removed from the statistics.  Uncertainties are calculated from small number statistics, following the method of \citet{gehrels86}. 
\begin{table}
\centering
\caption{SWIRE fields. In Col. 1 are the field names, in Col. 2 are the areas analysed, in Col. 3 are the RMS values calculated for $5\farcs25$ radii apertures (not aperture corrected).  Column 4 states the average number of sources per $1.75\arcmin$ radius circular cell.  The total number of  $1.75\arcmin$ radius circular fields analysed in SWIRE is 9981 (area $26.7$\,deg$^2$), totalling 72295 sources.}
\begin{tabular}{lcccc}
\hline \hline
Field Name & Area  & RMS & \={N}$_{sources}$ \\ 
\, & (deg$^2$) & ($\mu$Jy) & ($1.75\arcmin$ rad) \\
\multicolumn{1}{c}{(1)} & \multicolumn{1}{c}{(2)} & \multicolumn{1}{c}{(3)} & \multicolumn{1}{c}{(4)} \\ \hline
ELAIS N1 & 6.73  & 28.0 & 7.16 \\ 
ELAIS N2 & 4.60  & 28.3 & 7.83 \\ 
CDFS     & 7.69  & 27.9 & 7.16 \\ 
Lockman &  7.81 & 29.6 & 7.27 \\ 
XMM-LSS &  7.40  & 34.3 & 6.03 \\ \hline
\end{tabular}
\label{table1}
\end{table}
\begin{table*}
\caption{HzRG field Densities. HzRG names, redshifts, RMS noise values (for 5\farcs.25 radius apertures, not aperture corrected) and calculated background values in Cols. 1-4. Exposure time is stated in Col. 5. 500\,MHz radio luminosities quoted in Col. 6 are taken from \citet{debreuck10}.  $1.75\arcmin$ radius cells source counts and each field overdensity (with respect to the calculated SWIRE $1.75\arcmin$ radius fields Gaussian peak, as per Fig. \ref{figure:histogram}) along with their associated small-number statistical error in Cols. 7 and 8.  Column 9 shows the field overdensities as a function of $\sigma_{N}$, corresponding to the probability of finding that given number of sources in a blank, non targeted field. Fields classified as protocluster candidates (20) are in bold font.}
\begin{center}
\begin{tabular}{lcccccccc}
\hline \hline
  \multicolumn{1}{c}{HzRG} &
  \multicolumn{1}{c}{Redshift} &
  \multicolumn{1}{c}{RMS Noise} &
  \multicolumn{1}{c}{Background} &
  \multicolumn{1}{c}{Exposure} &
  \multicolumn{1}{c}{log$_{10}$(L$_{\rm{500MHz}}$)} &
  \multicolumn{1}{c}{\# Sources} &
  \multicolumn{1}{c}{$\delta\pm$Err} &
  \multicolumn{1}{c}{$\delta(\sigma)$} \\
$\,$ & $\,$ & [$\mu$Jy] & [mJy] & [s] &  [W\,Hz$^{-1}$] & ([$1.75\arcmin$ radius]) & [$N_{HzRG} /N_{SWIRE}$] & $\,$  \\
\multicolumn{1}{c}{(1)} & \multicolumn{1}{c}{(2)} & \multicolumn{1}{c}{(3)} & \multicolumn{1}{c}{(4)} & 
\multicolumn{1}{c}{(5)} & \multicolumn{1}{c}{(6)} & \multicolumn{1}{c}{(7)} & \multicolumn{1}{c}{(8)} & 
\multicolumn{1}{c}{(9)} \\ \hline
3C356.0	        & 1.079 &  19.5 & 0.76 &  450 & 28.35  & 8 & $1.5_{-0.5}^{+0.7}$ & 1.1 \\
MRC0037-258     & 1.100 &  22.1 & 0.75 &  900 & 27.72  & 7 & $1.3_{-0.5}^{+0.7}$ & 0.7 \\
3C368.0         & 1.132 &  27.6 & 0.73 &  450 & 28.52 & 11 & $2.1_{-0.6}^{+0.8}$ & 2.3 \\
\textbf{6C0058+495}   & 1.173 &  24.1 & 0.85 &  450 & 27.33 & 17 & $3.2_{-0.8}^{+1.0}$ & \textbf{4.7} \\
\textbf{3C65}  & 1.176 & 24.0 & 0.89 &  900 & 28.63 & 15 & $2.8_{-0.7}^{+0.9}$ & \textbf{3.9} \\
\textbf{3C266} & 1.275 & 17.3 & 1.07 &  900 & 28.54 & 15 & $2.8_{-0.7}^{+0.9}$ & \textbf{3.9} \\
MRC0211-256    & 1.300 & 20.6 & 0.79 &  900 & 27.78 & 12 & $2.3_{-0.6}^{+0.9}$ & 2.7 \\
\textbf{MRC0114-211} & 1.410 & 18.8 & 0.79 & 900 & 28.66 & 16 & $3.0_{-0.7}^{+1.0}$ & \textbf{4.3} \\
\textbf{LBDS53w069}   & 1.432 &  15.8 & 0.73 & 2100 & 26.30 & 20 & $3.8_{-0.8}^{+1.0}$ & \textbf{5.8} \\
\textbf{7C1756+6520}  & 1.480 &  18.7 & 0.78 &  450 & 27.40 & 15 & $2.8_{-0.7}^{+0.9}$ & \textbf{3.9} \\
7C1751+6809           & 1.540 &  18.0 & 0.78 &  450 & 27.46 & 9  & $1.7_{-0.6}^{+0.8}$ & 1.5 \\
LBDS53w091            & 1.552 &  12.6 & 0.73 & 2100 & 27.04 & 10 & $1.9_{-0.6}^{+0.8}$ & 1.9 \\
\textbf{3C470}        & 1.653 &  25.4 & 0.83 &  450 & 28.79 & 16 & $3.0_{-0.7}^{+1.0}$ & \textbf{4.3} \\
MRC2224-273           & 1.679 &  27.6 & 0.75 &  900 & 27.52 & 11 & $2.1_{-0.6}^{+0.8}$ & 2.3 \\
6C0132+330            & 1.710 &  28.3 & 1.18 &  900 & 27.64 & 9  & $1.7_{-0.6}^{+0.8}$ & 1.5 \\
\textbf{MRC1017-220}  & 1.768 &  23.0 & 0.72 &  900 & 27.94 & 15 & $2.8_{-0.7}^{+0.9}$ & \textbf{3.9} \\
3C239                 & 1.781 &  26.7 & 1.07 &  900 & 29.00 & 10 & $1.9_{-0.6}^{+0.8}$ & 1.9 \\
3C294                 & 1.786 &  25.2 & 0.49 &  900 & 28.96 & 9  & $1.7_{-0.6}^{+0.8}$ & 1.5 \\
\textbf{7C1805+6332}  & 1.840 &  18.5 & 0.78 &  450 & 27.78 & 21 & $4.0_{-0.9}^{+1.1}$ & \textbf{6.2} \\
6CE0820+3642          & 1.860 &  23.3 & 0.78 &  900 & 28.28 & 7 & $1.3_{-0.5}^{+0.7}$ & 0.7 \\
6CE0905+3955          & 1.883 &  17.5 & 0.78 &  900 & 28.17 & 8 & $1.5_{-0.5}^{+0.7}$ & 1.1 \\
MRC0324-228           & 1.894 &  16.2 & 0.79 &  900 & 28.49 & 7 & $1.3_{-0.5}^{+0.7}$ & 0.7 \\
\textbf{MRC0350-279}  & 1.900 &  27.0 & 0.78 & 450 & 28.25 & 13 & $2.5_{-0.7}^{+0.9}$ & \textbf{3.1} \\
MRC0152-209           & 1.920 &  20.6 & 0.79 &  360 & 28.20 & 12 & $2.3_{-0.6}^{+0.9}$ & 2.7 \\
MRC2048-272           & 2.060 &  23.1 & 0.76 &  900 & 28.72 & 11 & $2.1_{-0.6}^{+0.8}$ & 2.3 \\
\textbf{PKS1138-262}  & 2.156 &  9.2  & 1.08 & 9300 & 29.07   & 16 & $3.0_{-0.7}^{+1.0}$ &\textbf{4.3} \\

\textbf{4C40.36}      & 2.265 &  22.3 & 0.77 &  450 & 28.79   & 13 & $2.5_{-0.7}^{+0.9}$ & \textbf{3.1} \\
TXS0211-122           & 2.340 &  25.2 & 0.53 &  900 & 28.48 &  8 & $1.5_{-0.5}^{+0.7}$ & 1.1 \\
USS1707+105           & 2.349 &  17.7 & 0.79 &  900 & 28.63 & 10 & $1.9_{-0.6}^{+0.8}$ & 1.9 \\
USS1410-001           & 2.363 &  33.7 & 1.24 &  900 & 28.41 & 9  & $1.7_{-0.6}^{+0.8}$ & 1.5 \\
\textbf{LBDS53w002}   & 2.393 &  22.8 & 0.66 &  450 & 27.78 & 14 & $2.6_{-0.7}^{+0.9}$ & \textbf{3.5} \\
6C0930+389            & 2.395 &  23.5 & 1.39 &  900 & 28.41 & 7  & $1.3_{-0.5}^{+0.7}$ & 0.7 \\
MRC0406-244           & 2.427 &  28.7 & 0.76 &  450 & 29.03 & 9  & $1.7_{-0.6}^{+0.8}$ & 1.5 \\
\textbf{4C23.56}      & 2.483 &  32.3 & 0.77 &  450 & 28.93 & 17 & $3.2_{-0.8}^{+1.0}$ & \textbf{4.7} \\
MRC2104-242           & 2.491 &  28.4 & 0.78 &  900 & 28.84 & 12 & $2.3_{-0.6}^{+0.9}$ & 2.7 \\
WNJ1115+5016          & 2.540 &  19.2 & 1.47 &  900 & 27.82 & 8  & $1.5_{-0.5}^{+0.7}$ & 1.1 \\
\textbf{PKS0529-549}  & 2.575 &  30.5 & 0.79 &  450 & 29.16 & 18 & $3.4_{-0.8}^{+1.0}$ & \textbf{5.1} \\
\textbf{MRC2025-218}  & 2.630 &  25.9 & 0.76 &  900 & 28.74 & 27 & $5.1_{-1.0}^{+1.2}$ &\textbf{8.7} \\
\textbf{USS2202+128}  & 2.706 &  21.4 & 0.98 &  900 & 28.54 & 13 & $2.5_{-0.7}^{+0.9}$ & \textbf{3.1} \\
MG1019+0534           & 2.765 &  22.3 & 1.44 &  900 & 28.57 & 10 & $1.9_{-0.6}^{+0.8}$ & 1.9 \\
4C24.28               & 2.879 &  20.4 & 0.76 &  900 & 29.05 & 10 & $1.9_{-0.6}^{+0.8}$ & 1.9 \\
4C28.58               & 2.891 &  22.3 & 0.76 &  900 & 28.91 & 10 & $1.9_{-0.6}^{+0.8}$ & 1.9 \\
USS0943-242           & 2.923 &  17.3 & 0.66 &  450 & 28.62 & 9  & $1.7_{-0.6}^{+0.8}$ & 1.5 \\
\textbf{WNJ0747+3654} & 2.992 &  31.5 & 0.81 &  900 & 28.14   & 15 & $2.8_{-0.7}^{+0.9}$ &  \textbf{3.9} \\
B3J2330+3927          & 3.086 &  19.4 & 0.69 &  450 & 28.33 & 11 & $2.1_{-0.6}^{+0.8}$ & 2.3 \\
\textbf{MRC0316-257}  & 3.130 &  23.1 & 0.79 &  900 & 28.95 & 16 & $3.0_{-0.7}^{+1.0}$ & \textbf{4.3} \\
MRC0251-273           & 3.160 &  15.9 & 0.79 &  900 & 28.54 & 4 & $0.8_{-0.4}^{+0.6}$ & -0.5 \\
WNJ1123+3141          & 3.217 &  25.9 & 1.45 &  900 & 28.51 & 4 & $0.8_{-0.4}^{+0.6}$ & -0.5 \\
6C1232+39             & 3.220 &  19.2 & 1.10 &  900 & 28.93 & 8 & $1.5_{-0.5}^{+0.7}$ &  1.1 \\
TNJ0205+2242          & 3.506 &  26.5 & 0.77 &  900 & 28.46 & 6 & $1.1_{-0.4}^{+0.7}$ &  0.3 \\
TNJ0121+1320          & 3.516 &  28.3 & 0.78 &  900 & 28.49 & 5 & $0.9_{-0.4}^{+0.6}$ & -0.1 \\
TXJ1908+7220          & 3.530 &  26.0 & 0.48 & 450 & 29.12 & 12 & $2.3_{-0.6}^{+0.9}$ &  2.7 \\
USS1243+036           & 3.570 &  29.4 & 1.45 & 900 & 29.23 & 12 & $2.3_{-0.6}^{+0.9}$ &  2.7 \\
WNJ1911+6342          & 3.590 &  19.4 & 0.77 & 450 & 28.14 & 7  & $1.3_{-0.5}^{+0.7}$ &  0.7 \\
MG2144+1928           & 3.592 &  25.6 & 0.85 & 900 & 29.08 & 12 & $2.3_{-0.6}^{+0.9}$ &  2.7 \\
6C0032+412            & 3.670 &  21.4 & 0.77 & 450 & 28.75 & 11 & $2.1_{-0.6}^{+0.8}$ &  2.3 \\
4C60.07               & 3.788 &  31.4 & 0.77 & 450 & 29.20 & 9  & $1.7_{-0.6}^{+0.8}$ &  1.5 \\
4C41.17               & 3.792 &  23.1 & 0.78 & 900 & 29.18 & 11 & $2.1_{-0.6}^{+0.8}$ &  2.3 \\
TNJ2007-1316          & 3.840 &  18.7 & 0.76 & 900 & 29.13 & 9  & $1.7_{-0.6}^{+0.8}$ &  1.5 \\
TNJ1338-1942          & 4.110 &  32.0 & 0.78 & 900 & 28.71 & 10 & $1.9_{-0.6}^{+0.8}$ &  1.9 \\
8C1435+635            & 4.250 &  21.0 & 0.76 & 450 & 29.40 & 9  & $1.7_{-0.6}^{+0.8}$ &  1.5 \\
\textbf{6C0140+326}   & 4.413 &  24.3 & 1.39 & 900 & 28.73 & 18 & $3.4_{-0.8}^{+1.0}$ &  \textbf{5.1} \\
TNJ0924-2201          & 5.195 &  23.1 & 0.74 & 450 & 29.51 & 7 & $1.3_{-0.5}^{+0.7}$ & 0.7 \\
\hline
\end{tabular}
\label{table2}
\end{center}
\end{table*}
\section{Analysis}
\subsection{Comparison with SWIRE data}
\begin{center}
\begin{figure*}
\centering
\includegraphics[width=0.8\linewidth,angle=270]{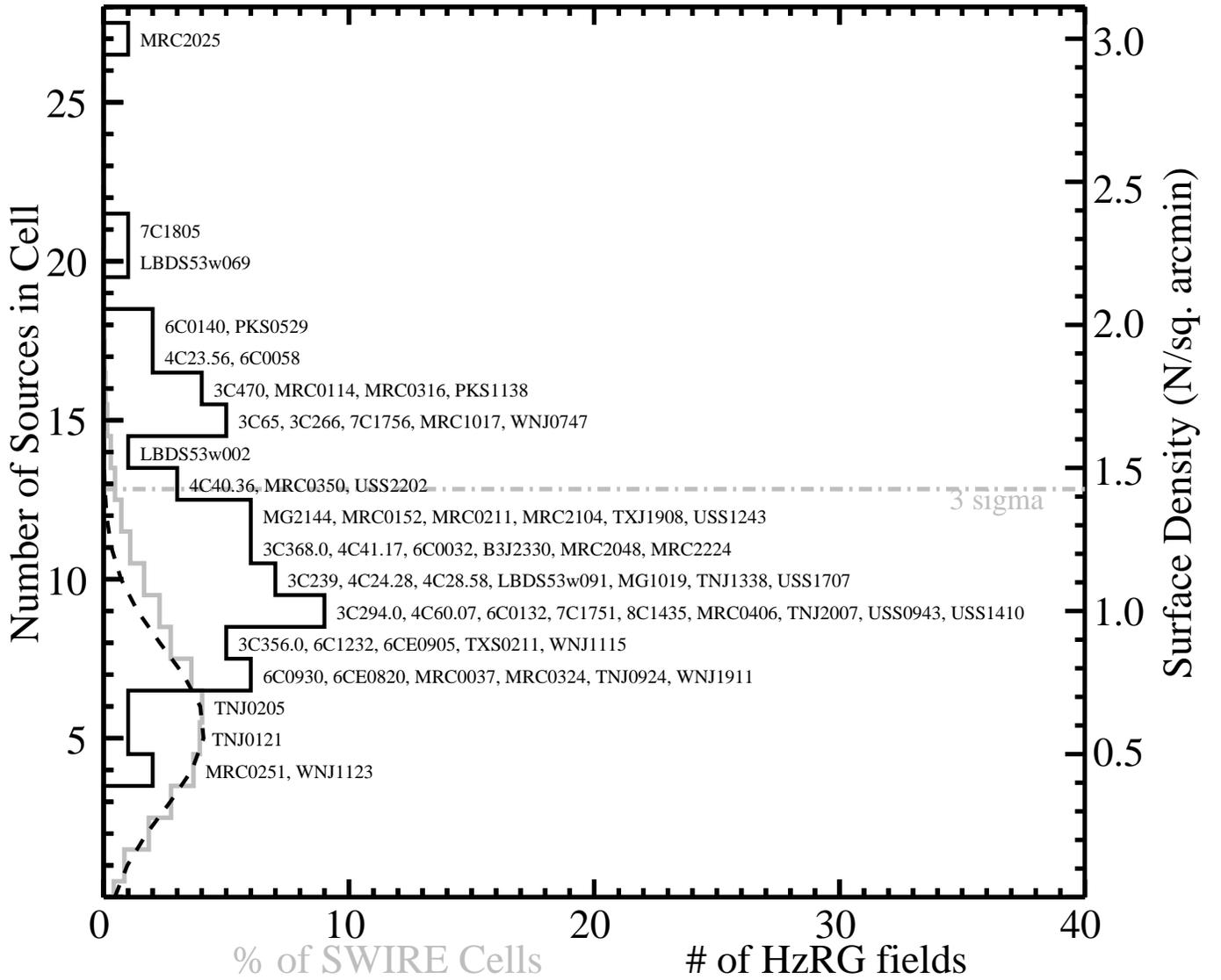}
\caption{A comparison of HzRG source counts with the reference SWIRE fields. The distribution of the SWIRE 24\,\um\, source counts in $\sim 10,000$ cirular cells of $1.75\arcmin$ radius, representing five SWIRE fields, is plotted as a grey histogram.  A Gaussian fit to the lower part of this distribution is overlayed as a dashed line. A $3\sigma$ above the SWIRE mean is indicated as a dot-dash line.  A histogram of HzRG field densities is overlayed along with each HzRG name.  The corresponding source number density is indicated on the right axis.}
\label{figure:histogram}
\end{figure*}
\end{center}

Figure \ref{figure:histogram} shows the source density distribution of the five SWIRE fields.  The grey histogram represents the percentage of cells containing a given number of sources.  The dashed line shows the Gaussian fit to the lower portion of this distribution, and exemplifies the positive skew in the SWIRE data due to clustering.  Our values for the Gaussian fit are  $\langle{N}\rangle = 5.28$ with $\sigma_N = 2.52$.  
Superimposed upon this plot is a histogram of the HzRG field densities, whereby the histogram height represents the number of HzRG fields for a given number of sources. Along with the HzRG histogram the HzRG field names are included.  We also overplot our criterion for protocluster candidacy of $N_{HzRG} \ge \langle{N}\rangle + 3\sigma_N = 12.84$, a cut which only 6\% of SWIRE cells exceed.  This comparison of our HzRG field data with the SWIRE fields shows such an overdensity in a significant minority (20 [32\%]) of HzRG fields. The mean density of the HzRG fields is $11.6\pm6.3$, translating to a mean overdensity ($\delta = \frac{\bar{N}_{HzRG}}{\bar{N}_{SWIRE}}$) of $2.2\pm1.2$ over the reference SWIRE data, with a Kolmolgorov-Smirnov probability that the two datasets are drawn from the same distribution of $1.2 \times 10^{-12}$.  Individual field source densities are included in Table \ref{table2} 
\subsection{Comparison with IRAC}
\begin{center}
\begin{figure}
\centering
\includegraphics[width=0.95\linewidth]{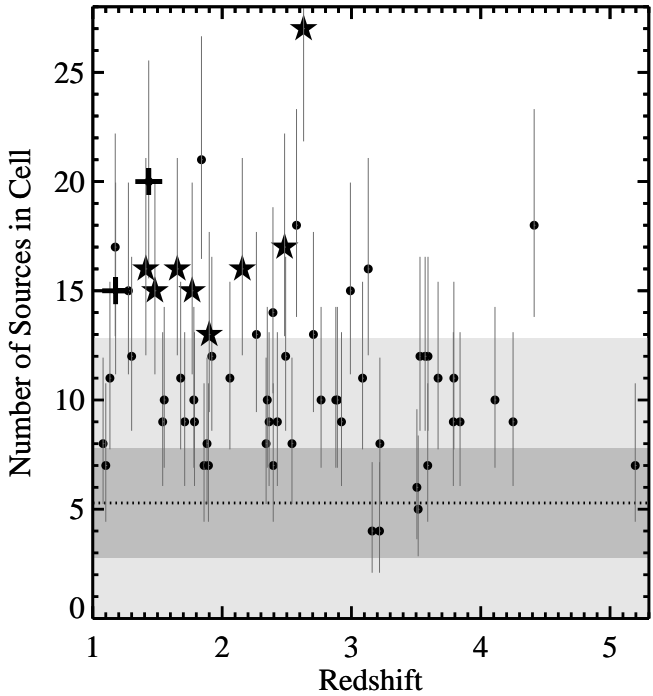}
\caption{Source counts around each HzRG versus HzRG redshift.  The dotted line indicates the SWIRE mean source count for a $1.75\arcmin$ radius cell, the dark shaded region denotes $\pm1\sigma$ and the light shaded region denotes  $\pm3\sigma$ on the SWIRE data. Starred symbols denote those sources which are considered candidate clusters in G12 (8) while the plus symboled sources (2) are those which are not studied in G12.  Note that $z > 3$ sources are not deep enough to clarify protocluster candidacy in G12.  The uncertainties on the source counts are derived from small number statistics.}
\label{figure:ODvsz}
\end{figure}
\end{center}

G12 study the environments of $1.2 \le z \le 3.0$ HzRGs and compare the densities of sources with IRAC $[3.6]-[4.5]>-0.1$ (AB) (within a certain flux density cut) found within one arcminute of the HzRG.  The imposed colour cut selects against galaxies below $z\sim1.2$, but does not distintinguish between passive and active galaxies \citep{papovich08}.  Comparing the main result of this paper, a significant fraction of our sources confirm the protocluster candidacies proposed in G12.  The combined result of this comparison is shown in Fig. \ref{figure:ODvsz}, where the 24\um\, source number in $1.75\arcmin$ radius circular cells is plotted against radio galaxy redshift.  The dotted line, dark shaded region and light shaded region represent the SWIRE mean density, $\langle N \rangle$, $1\sigma_N$ and $3\sigma_N$ deviations, all extracted from the Gaussian fit described above.  Plotted star symbols represent those sources with an overdensity in our work of $\delta_{MIPS} \ge \langle N \rangle + 3\sigma_N$ and in G12 with an overdensity of  $\delta_{IRAC} \ge \langle N \rangle + 2\sigma_N$. Plotted circles represent those remaining sources present in both catalogs, and the plotted plus symbols represent the fields which are not analysed in G12: \textbf{3C65} (z=1.176) and \textbf{LBDS53W059} (z=1.432).  Notes on protocluster candidates are included in Sec. \ref{individ} and shown in Fig. \ref{fig:cutouts}.

\begin{center}
\begin{figure}
\centering
\includegraphics[width=0.95\linewidth]{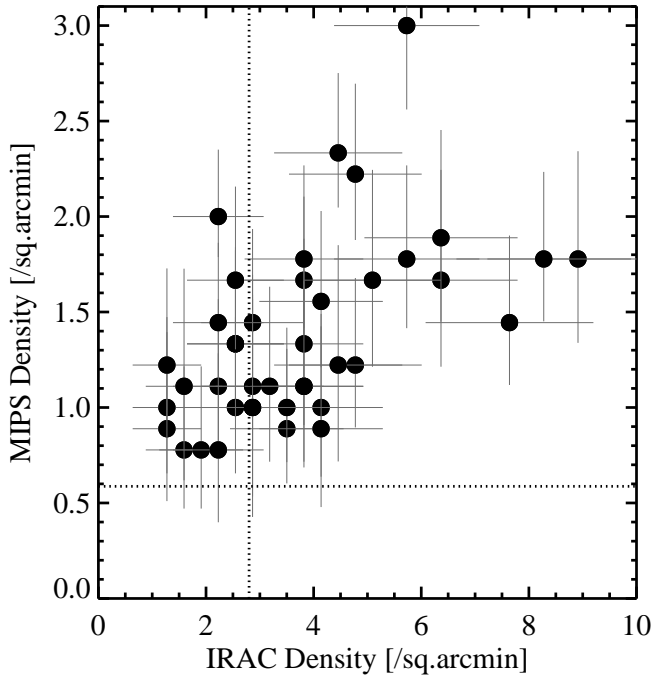}
\caption{Comparison of MIPS field densities with IRAC $[3.6]-[4.5]>-0.1$ field densities from G12 (\textbf{$1.2 \le z \le 3.0$}).  The horizontal and vertical dotted line correspond to the mean SWIRE 24\um\, and IRAC colour-selected values from SWIRE, respectively.  A relatively strong correlation between source densities is found.}
\label{figure:ODvsIRAC}
\end{figure}
\end{center}

\begin{center}
\begin{figure}
\centering
\includegraphics[width=0.95\linewidth]{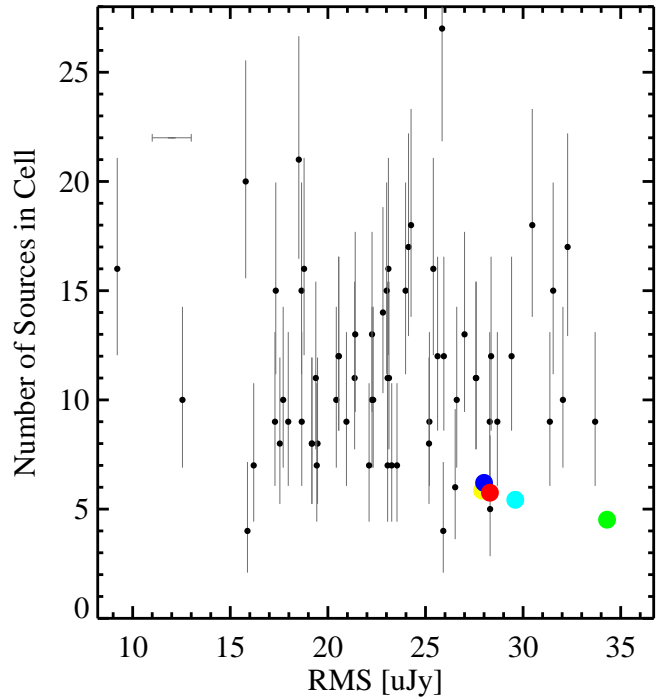}
\caption{HzRG field density versus RMS noise. No obvious correlation between the two values is found.  The error bar plotted in the upper left in included to give an indication of a typical uncertainty in the RMS values.  SWIRE field values are plotted as larger circles where yellow, blue, red, green and cyan correspond to CDFS, EN1, EN2, XMM-LSS and Lockman fields respectively.}
\label{figure:NumbervsRMS}
\end{figure}
\end{center}

Figure \ref{figure:ODvsIRAC} shows the source field densities as measured in IRAC by G12 versus the $24$\,\um\, source densities measured here.  A strong correlation between the two values is found, despite the wavebands probing different populations. Of particular interest for further study may be those environments which are found to be significantly dense in one waveband but unremarkable or even underdense in the other, as these field abnormalities may enlighten us with regard to structure formation. The Spearman Rank correlation is $\rho = 0.577$, implying a statistically significant correlation (\eg there is a $99.992$\% probability that the two parameters are correlated).
As a check of the quality of our data, the RMS noise is plotted against the $24$\,\um\, source density (Fig. \ref{figure:NumbervsRMS}).  We expect there to be no obvious correlations between these values, and indeed no correlation is found.  The Spearman Rank Correlation coefficient for these data is $\rho=0.08$, implying only a $48.6$\% probability of a correlation between the values.
\subsection{Evolution with Radio Luminosity}
\begin{center}
\begin{figure}
\centering
\includegraphics[width=0.95\linewidth]{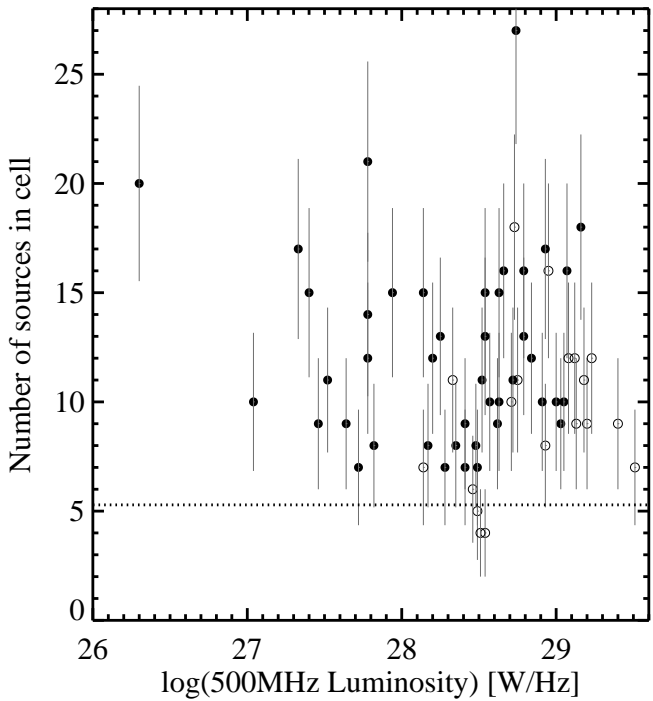}
\caption{HzRG $24$\,\um\, field densities versus HzRG 500\,MHz luminosities \citep[taken from ][]{debreuck10}. The HzRGs with $z>3.0$ are plotted as open symbols.}
\label{figure:radio}
\end{figure}
\end{center}

Plotting of $24$\,\um\, source density versus HzRG 500MHz luminosity (Fig. \ref{figure:radio}) shows no significant correlation between the two parameters.  Though no work in the literature has had a large enough sample to statically prove or disprove any correlation, suggestions that high luminosity radio galaxies lie within a more dense medium have been put forward \citep{venemans07,miley08,falder10}.  A possible explanation for this is that higher luminosity radio galaxies lie in a medium with a higher than average density, thus being more conducive to cluster formation.  Our work, alongside that of G12 are the first capable of making such a statistically significant comparison and we find no such correlation between source density and HzRG $500$\,MHz luminosity.  The Spearman Rank Correlation coefficient recovered for these data is $\rho=0.04$, implying only a $25$\% probability that the two parameters are correlated. 
We note that there is a small redshift-radio power degeneracy in our sample (especially at $z>3$, see Fig. \ref{figure:radio}).  Given that redshift and density are clearly correlated (see Fig \ref{figure:ODvsz}) this may imprint a correlation between source density and radio power.  We take this into account by using the Spearman Partial Rank correlation coefficient \citep{mac82} which reduces the correlation coefficient to $\rho=0.01$.  We conclude that the 24\,\um\, source density does not depend on radio power.

\subsection{K-correction and Population Type}
\label{AGN}
\begin{center}
\begin{figure}
\centering
\includegraphics[width=0.95\linewidth]{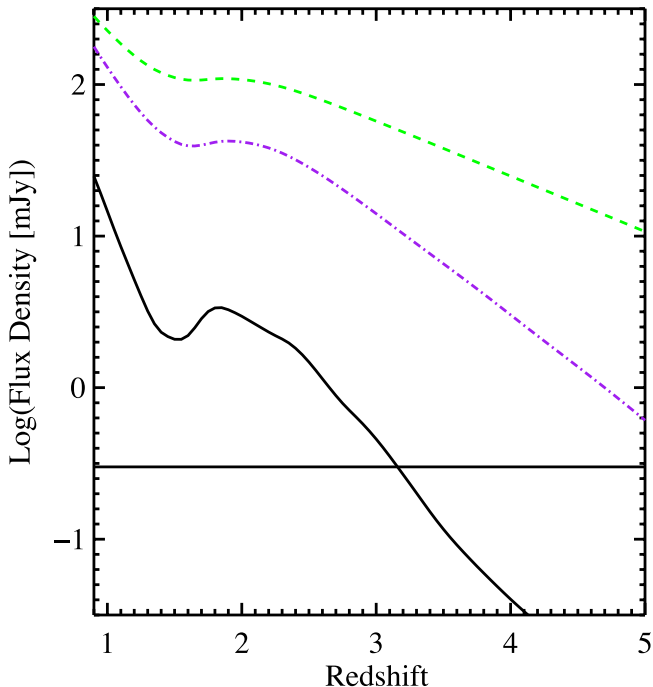}
\caption{Observed MIPS 24\,\um\, flux density versus redshift.  Solid black line represents a starburst galaxy with a star formation rate of 1000\,M$_\odot\,$yr$^{-1}$. The dashed green line and dot-dash blue line represent a face-on and an edge-on quasar with $L_{24\um} = 10^{45}$\,erg s$^{-1}$. The horizontal line represents the $0.3$\,mJy flux density cut.}
\label{figure:SED}
\end{figure}
\end{center}

Our sample covers a wide range in redshifts, and the observed $24$\,\um\, band therefore covers significantly different restframe wavelengths.  To examine these K-correction effects, we map the Spectral Energy Distributions (SEDs) of a starburst galaxy \citep{siebenmorgen2007} as well as face- and edge-on quasars \citep{pier92} through the MIPS $24$\,\um\, transmission window for redshifts $1 < z <  5.2$ (Fig. \ref{figure:SED}).

It is interesting to note a number of things; (1) the relative flatness of the quasar flux density over the entire redshift range, due to hot dust entering the MIPS transmission window; and (2) the bump in the starburst SED at $z \sim 1.5$, due to the PAH emission lines (dominated by 7.7\,\um) entering the MIPS transmission window.  Note that the smoothness of the starburst line is due to the broad MIPS 24\,\um\, transmission window covering some 9\,\um\, between $\sim$20-29\,\um, thus smoothing out the more narrow PAH features.  Also note that even for such a high SFR starburst, galaxies fall below the flux density cut beyond redshift three. Finally, the orientation of quasars will have a large impact on their detectability at lower luminosities.  

By comparing a high SFR starburst to a moderately luminous quasar, as shown in Fig. \ref{figure:SED}, we can make the assumption\footnote{if we assume no evolution of the luminosity functions.} that above a redshift of $z \sim 3$ we are probing almost exclusively the quasar population.  On the other hand, below $z \sim 3$, we are simultaneously probing the quasar and starburst populations, with starbursts become ever-more important at redshifts two and below.
We are likely to see higher densities in fields at lower redshifts.  This is indeed what we observe; as shown in Fig. \ref{figure:ODvsz} the \emph{highest} density fields ($\delta(\sigma) \gtrsim 4.5$) almost exclusively lie below a redshift of three. \citet[see Fig. 2]{reddy06} compare spectroscopically confirmed $24$\,\um\, sources in the GOODS-North field between redshifts $1 \le z \le 2.6$.  For a flux density cut of $f_{24\um} \ge 0.3$\,mJy, they find only Sub-Millimetre Galaxies (SMGs) and X-ray sources in the field, consistent with probing the starburst galaxy and AGN populations.

The minimum AGN luminosities which we can probe with our imposed flux density cut are $3.4\times10^{41}$\,erg s$^{-1}$ at $z=1$ to $1.9\times10^{43}$\,erg s$^{-1}$ at $z=5.2$.  Compare this with the minimum SFRs which can equivalently be probed; SFR$=100$\,M$_\odot\,$yr$^{-1}$ at $z=1$ to a SFR $\sim$1000\,M$_\odot\,$yr$^{-1}$ at $z=3$. These values have been calculated using the relevant formulae from \citet{rieke09}.

Since we do not know the physical nature of every object in each frame, let alone the redshift of each of these sources, it is not possible to perform K-corrections.  For K-corrections to be calculated we will need to scrutinise the complimentary IRAC data for these fields, as well as gain data in the Far-IR.  This process is currently being undertaken and will be the focus of future works. However, it is beyond the scope of this paper, which concentrates only on the statistical properties of HzRG fields.
\subsection{Evolution with Redshift}
\label{sect:analysis:RMS}
Figure \ref{figure:ODvsz} also shows the HzRG field densities versus HzRG redshift.  If \textbf{6C0140+326} (z=4.413) is a lensed system (as suggested in \citealt{lacy99}) and as such possibly contaminated by artificially brightened sources, then this field should be removed from our sample.  If we do remove this field, then  we begin to see a general trend towards smaller-overdensities at early epochs.  This could be interpreted in two ways: (1) the cluster luminosity function is truly evolving over time; or (2) the imposed flux density cut is too high to probe comparable populations at higher redshifts, as discussed above. 

Since we are probing different populations at lower ($z=1-2$) and higher ($z=3-5$) redshifts, with a transitory region between them, higher source densities at later epochs is the behavior that one would expect. However, since we also expect structure formation to be evolutionary, and not instantaneous, observations of higher overdensities at lower redshifts cannot be assigned to selection bias alone. The mid-IR luminosity function has been pushed out to $z\sim2.5$ through stacking analysis \citep{caputi07,rod10} though is essentially unconstrained at $z>2.5$.  We do not expect a significant increase in number density at higher redshifts as this would be inconsistent with the star-formation rate density-evolutionary plot \citep{lilly96,madau96}; the high-end of the mid-IR luminosity function must drop off at high redshift, though constraints are too poor to infer population statistics.

\subsection{Notes on Individual Sources}
\label{individ}
\begin{figure*}
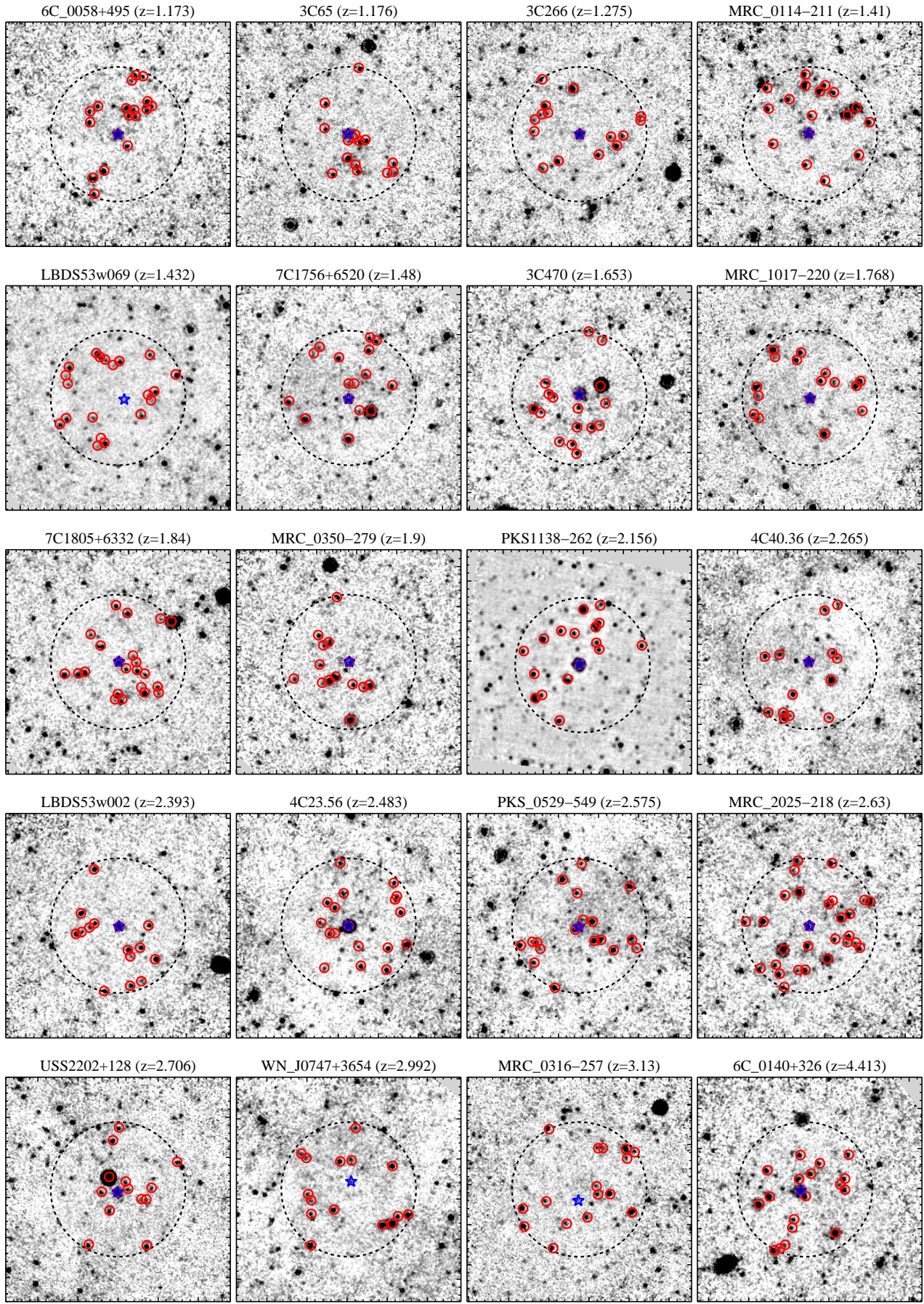

\centering
\subfigure{\includegraphics[width=0.225\linewidth]{6C_0058+495_cutout_circ.epsi}}
\subfigure{\includegraphics[width=0.225\linewidth]{3C65_cutout_circ.epsi}}
\subfigure{\includegraphics[width=0.225\linewidth]{3C266_cutout_circ.epsi}}
\subfigure{\includegraphics[width=0.225\linewidth]{MRC_0114-211_cutout_circ.epsi}} \\
\subfigure{\includegraphics[width=0.225\linewidth]{LBDS53w069_cutout_circ.epsi}}
\subfigure{\includegraphics[width=0.225\linewidth]{7C1756+6520_cutout_circ.epsi}}
\subfigure{\includegraphics[width=0.225\linewidth]{3C470_cutout_circ.epsi}}
\subfigure{\includegraphics[width=0.225\linewidth]{MRC_1017-220_cutout_circ.epsi}} \\
\subfigure{\includegraphics[width=0.225\linewidth]{7C1805+6332_cutout_circ.epsi}}
\subfigure{\includegraphics[width=0.225\linewidth]{MRC_0350-279_cutout_circ.epsi}}
\subfigure{\includegraphics[width=0.225\linewidth]{PKS1138-262_cutout_circ.epsi}}
\subfigure{\includegraphics[width=0.225\linewidth]{4C40.36_cutout_circ.epsi}} \\
\subfigure{\includegraphics[width=0.225\linewidth]{LBDS53w002_cutout_circ.epsi}}
\subfigure{\includegraphics[width=0.225\linewidth]{4C23.56_cutout_circ.epsi}}
\subfigure{\includegraphics[width=0.225\linewidth]{PKS_0529-549_cutout_circ.epsi}}
\subfigure{\includegraphics[width=0.225\linewidth]{MRC_2025-218_cutout_circ.epsi}} \\
\subfigure{\includegraphics[width=0.225\linewidth]{USS2202+128_cutout_circ.epsi}}
\subfigure{\includegraphics[width=0.225\linewidth]{WN_J0747+3654_cutout_circ.epsi}}
\subfigure{\includegraphics[width=0.225\linewidth]{MRC_0316-257_cutout_circ.epsi}}
\subfigure{\includegraphics[width=0.225\linewidth]{6C_0140+326_cutout_circ.epsi}}
\caption[]{\tiny{Cutouts of all HzRG $24\,\um$\, fields with a $3\sigma$ significance source overdensity. The dotted circle represents the  $1.75\arcmin$ radius circular sub-field that is analysed further, and plotted in red are the source for which $f_{24\mu m} \ge 0.3$\,mJy.  All fields are scaled to the same  $5.8'\times5.8'$ angular size, with North/East facing up/left. The HzRG is plotted as a blue star in each frame.  Note that \textbf{PKS1138-262} is much deeper GTO data.} A colour counterpart is available in the online version of this article.}
\label{fig:cutouts}
\end{figure*}

For the sake of a well defined sample, we only select $24$\,\um\, targets with an $N_{\rm HzRG} > \langle$N$_{\rm SWIRE}\rangle +3\sigma$ as cluster candidates.  This accounts for 20 (32\%) of our sample. For all sources that meet this criterion, cutouts are included in Fig. \ref{fig:cutouts}. Fields are sorted by increasing redshift. Quoted sigma values correspond to the probability of finding protocluster candidates compared to a blank, non-targeted field.
\subsubsection{\textbf{6C0058+495} (z=1.173)} 
We find a 24\,\um\, source overdensity of $4.7\sigma$, with the sources in this field having a clumpy structure.  This field is our lowest redshift protocluster candidate.
\subsubsection{\textbf{3C65} (z=1.176)}
This field was first studied by \citet{best00}, who found a $K$-band overdensity.  Our study confirms an overdensity in 3C65 at a $3.9\sigma$ level.
\subsubsection{\textbf{3C266} (z=1.275)} 
This field has also been studied in G12, but its source overdensity was not significant enough to declare protocluster candidacy. This work however finds a 24\,\um\, source overdensity of $3.9\sigma$, making it a firm candidate.
\subsubsection{\textbf{MRC0114-211} (z=1.41)}
With an overdensity of IRAC sources significant at a $6\sigma$ level compared to the field, MRC0114-211 is the densest field and most promising cluster candidate in G12.  We also find an overdensity of 24\,\um\, sources at the 4.3$\sigma$ significance.
\subsubsection{\textbf{LBDS53W069} (z=1.432)}
This is formally not a radio galaxy according to the radio luminosity cutoff defined by \citet{seymour2007}. Nevertheless, our work shows a significant 24\,\um\, source overdensity at $5.8\sigma$; observationally this field is very dense with no preferred spatial segregation.  This field was not included in G12's $[3.6]-[4.5] > -0.1$ analysis, but it has been calculated that this field  would represent a $4.8\sigma$ overdensity.
\subsubsection{\textbf{7C1756+6520} (z=1.48)}
This field is a known protocluster, first identified in \citet{galametz09} and spectroscopically confirmed in \citet{galametz10}. This work confirms that 7C1756+6520 is also a protocluster candidate at $24\,\um$, with a source overdensity of $3.9\sigma$.
\subsubsection{\textbf{3C470} (z=1.653)}
This field was first studied in G12, with a source overdensity of $2.9\sigma$.  Our work quotes a $24\,\um$ source overdensity of $4.3\sigma$.
\subsubsection{\textbf{MRC1017-220} (z=1.768)}
This field was first studied by \citet{cimatti00}, who found an overdensity of EROs.  Studies of near-IR selected galaxies \citep{galametz10} and IRAC-selected galaxies (G12), however, did not find a significant overdensity.  Our work finds a $24\,\um$ source overdensity at the $3.9\sigma$ level.
\subsubsection{\textbf{7C1805+6332} (z=1.84)}
This field was first studied in \citet{galametzPhD}, where no near-IR overdensity was reported. This field was also studied in G12, which again did not find a significant overdensity.  This work finds a $24\,\um$ source overdensity at $6.2\sigma$, making it a very strong protocluster candidate. Observationally, sources lie in a filamentary structure in the NE-SW direction. 
\subsection{\textbf{MRC0350-279} (z=1.9)}
Field studied in G12, with an IRAC source overdensity of $\sim 5.2\sigma$ derived.  Here, the 24\,\um\, source overdensity is 3.1$\sigma$.  Observationally the field is very much clustered towards the East of the HzRG.
\subsubsection{\textbf{PKS1138-262} (z=2.156)}
\textbf{PKS1138-262} is a well studied protocluster environment with overdensities of Ly$\alpha$ and X-ray sources \citep{pentericci00,pentericci02}, as well as H$\alpha$ emitters and EROs \citep{kurk04b,kurk04a}. This work finds an overdensity of $24\,\um$ sources at $4.3\sigma$. 
\subsubsection{\textbf{4C40.36} (z=2.265)}
This field was studied in G12, who found an underdensity of red IRAC sources.  Our work quotes a $24\,\um$ source overdensity of $3.1\sigma$. Sources appear to lie in a filamentary E-W structure.
\subsubsection{\textbf{LBDS53W002} (z=2.393)}
This field was studied in G12, but no significant overdensity was found.  The $24\,\um$ source overdensity quoted here is $3.5\sigma$. The structure appears to have some form of preferred direction, though this remains ambiguous without a larger field of view.
\subsubsection{\textbf{4C23.56} (z=2.483)}
There are many works covering this field, beginning with \citet{knopp97} that quote a ``marginally significant excess of objects'' in the $K$-band. Other works have found overdensities in the near-IR \citep{kajisawa06}, of H$\alpha$ emitters \citep{tanaka11} and mid-IR (G12).  This work finds a $24\,\um\,$ source overdensity at a $4.7\sigma$ level.
\subsubsection{\textbf{PKS0529-549} (z=2.575)}
This field was studied in G12 and its environment was found to be underdense.  Here, however, a significant $24\,\um\,$ source overdensity of $5.1\sigma$ is found. The sources appear to be very strongly clustered within the $1.75\arcmin$ radius frame, making this field a firm protocluster candidate at $24\,\um$.
\subsubsection{\textbf{MRC2025-218} (z=2.63)} 
This is our highest density field, and has recently been studied by G12, quoting a $2.9\sigma$ overdensity. Our result shows a $24$\,\um\, source overdensity at $8.6\sigma$, with 27 ($f_{24\um} > 0.3$\,mJy) sources in a $1.75\arcmin$ radius field. This exceeds the number of sources in every $1.75\arcmin $ radius SWIRE sub-field from our analysis.
\subsubsection{\textbf{USS2202+128} (z=2.706)}
This field was studied in G12, who found no overdensity. Here, the $24\,\um\,$ source overdensity is quoted as $3.1\sigma$ making it a firm protocluster candidate. Clustering appears mostly in the North of the frame, centred on a bright object $\sim30''$ from the HzRG.
\subsubsection{\textbf{WNJ0747+3654} (z=2.992)}
An $24\,\um\,$ overdensity of $3.9\sigma$ is found in this field, making \textbf{WNJ0747+3654} a strong protocluster candidate. Upon visual inspection, local clustering of sources (within the $1.75\arcmin$ radius cell) appears to be taking place, though on larger scales (outside of the cell) no spatial segregation appears to be present.
\subsubsection{\textbf{MRC0316-257} (z=3.13)} 
This target is a known Ly$\alpha$ and \texttt{[OIII]} protocluster \citep{venemans05,venemans07,maschietto08, kuiper10}, which our work also confirms with a $24\,\um\,$ source overdensity of $4.3\sigma$.
\subsubsection{\textbf{6C0140+326} (z=4.413)}
This is a known protocluster, having recently been studied by \citet{kuiper11}, with a Ly$\alpha$ overdensity of $9\pm5$ over a random field distribution.  Our result confirms a 24\,\um\, source overdensity in the vicinity of 6C0140+326 at $5.1\sigma$. However, if it is lensed as suggested in \citealt{lacy99}, then this field may be contaminated by artificially brightened sources.
\section{Conclusions}
Mid-IR emission is a powerful tracer of both AGN (torus-dust) and starburst galaxies, making mid-IR observations ideal for finding galaxy clusters in the early Universe. Using targeted observations in the mid-IR, we have investigated the fields of 63 HzRGs and find a statistically significant source overdensity over SWIRE reference fields data.  We find a large minority of our targeted fields, 20, or 32\%, to be overdense to at least a $3\sigma$ significance, confirming 11 known protoclusters (or existing protocluster candidates).  We identify nine new protocluster candidates.  Targeted mid-IR observations of HzRG fields shows to be a powerful technique for protocluster candidate selection.

These overdensities are indicative of the presence of protoclusters with active and star$-$forming galaxies at high redshifts.  Our results indicate a redshift evolution of source density, but this may be due to an insensitivity to star-forming galaxies at high redshifts.  Our dataset is compared with recent results of IRAC colour selected densities and correlations between the data are found.

Further work regard these data will involve spectroscopic follow-ups for each of the cluster candidates in order to confirm members.  Meanwhile Far-IR photometry currently being undertaken, in conjunction with the existing \emph{Spitzer} data will allow a more complete knowledge of the population which make up these overdense fields.

\appendix
\section{Photometry Parameters}
\label{app2}
\subsection{Basic SExtractor Parameters}
\begin{itemize}
 \item CLEANing was turned off
 \item NPIXEL, the minimum number of adjacent pixels above the detection 
threshold to qualify a source, is set to 4.
 \item DEBLEND\_NTHRESH, the number of deblending sub-thresholds is set 
to 16.
 \item DEBLEND\_MINCONT, the minimum contrast parameter for deblending 
is set to 0.01.
 \item PHOT\_APERTURES, the fixed aperture \textbf{diameter} for photometry
 is set depending on whether the SWIRE frames of the HzRG frames are being analysed.  Since the re-sampling is different this value must be set to 8.75 or 8.4 respectively.
 \item PHOT\_AUTOPARAMS, The Kron$-$Factor and minimum Kron radius 
respectively are set to 2.55, 3.5.
 \item THRESH\_TYPE, is set to RELATIVE.  This constrains DETECT\_THRESH 
to be \emph{relative} to the background.
\item DETECT\_THRESH is set to 3.
 \item BACK\_SIZE, the background mesh size is set to 32 pixels.
 \item BACK\_FILTERSIZE, is set to 3.
 \item BACKPHOTO\_TYPE, is set to  LOCAL.
 \item SEEING\_FWHM, the Stellar FWHM is set to 5.0 (arcseconds).
 \item GAIN is set to 0.0 (e-/ADU)
 \item WEIGHT\_TYPE is set to MAP\_VAR, since the weight map has been edited
 to be an absolute variance map.
 \item WEIGHT\_GAIN is set to N
\end{itemize}

\begin{acknowledgements}
This work is based on observations made with the \emph{Spitzer Space Telescope}, which is operated by the Jet Propulsion Laboratory, California Institute of Technology under a contract with NASA.  JHM would like to thank the ESO DGDF for helping make this work possible.
\end{acknowledgements}

\end{document}